\documentclass[onecolumn,10pt]{asme2ej}

\usepackage{epsfig} 
\usepackage{setspace}

\usepackage{graphicx}
\usepackage{color}
\usepackage{pifont}
\usepackage{latexsym}
\usepackage{amssymb}
\usepackage[mathb,mathx]{mathabx}

\title{Study of the cavitating instability on a grooved Venturi profile}

\author{Am\'elie Danlos
    \affiliation{
	Post-Doctoral student\\
	DynFluid Laboratory, EA 92\\
	Arts et M\'etiers ParisTech\\
	151 Boulevard de l'H\^opital\\
	75013 Paris, France\\
    Email: amelie.danlos@ensam.eu
    }	
}

\author{Jean-Elie M\'ehal \\
    \affiliation{ PhD student\\
	DynFluid Laboratory, EA 92\\
	Arts et M\'etiers ParisTech\\
	151 Boulevard de l'H\^opital\\
	75013 Paris, France\\
    }
}

\author{Florent Ravelet \\
    \affiliation{ Associate Professor\\
	DynFluid Laboratory, EA 92\\
	Arts et M\'etiers ParisTech\\
	151 Boulevard de l'H\^opital\\
	75013 Paris, France\\
    Email: florent.ravelet@ensam.eu}
}

\author{Olivier Coutier-Delgosha \\
    \affiliation{Professor\\
	Laboratoire de M\'ecanique de Lille, UMR 8107\\
	Arts et M\'etiers ParisTech\\
	8 Boulevard Louis XIV\\
	59046 Lille, France\\
    Email: olivier.coutier@ensam.eu}
}

\author{Farid Bakir \\
	\affiliation{Professor\\
    DynFluid Laboratory, EA 92\\
	Arts et M\'etiers ParisTech\\
	151 Boulevard de l'H\^opital\\
	92290 Paris, France\\
    Email: farid.bakir@ensam.eu}
}

\begin{document}

\maketitle    

\doublespacing
\begin{abstract}
{\it Instabilities of a partial cavity developed on an hydrofoil, a converging-diverging step or in an inter-blade channel, have already been investigated in many previous works. The aim of this study is to evaluate a passive control method of the sheet cavity. According to operating conditions, cavitation can be described by two different regimes: an unstable regime with a cloud cavitation shedding and a stable regime with only a pulsating sheet cavity. Avoiding cloud cavitation can limit structure damages since this regime is less agressive. The surface condition of a converging-diverging step is here studied as a solution to control the cavitation regime. This study discusses the effect of longitudinal grooves, on the developed sheet cavity. Analyzes conducted with Laser Doppler Velocimetry, visualisations and pressure measurements show that the grooves geometry, and especially the groove depth, acts on the sheet cavity dynamics and can even suppress the cloud cavitation shedding.
}
\end{abstract}

\begin{nomenclature}
\entry{$C_{f}$}{Friction coefficient}
\entry{$H$}{Sheet cavity mean height}
\entry{$H^{*}$}{Undimensional sheet cavity mean height}
\entry{$H_{throat}$}{Height of the section at the Venturi throat}
\entry{$K$}{Karman constant}
\entry{$L$}{Sheet cavity mean length}
\entry{$L^{*}$}{Undimensional sheet cavity mean length}
\entry{$N$}{Number of grooves}
\entry{$P$}{Relative pressure}
\entry{$Q$}{Flow rate}
\entry{$Re_{y_{throat}}$}{Reynolds number}
\entry{$T$}{Water temperature}
\entry{$St_{L}$}{Strouhal number}
\entry{$d$}{Diameter blur of grooves}
\entry{$e$}{Width of the ridge}
\entry{$f$}{Shedding frequency}
\entry{$f_{L}$}{Frequency of the sheet cavity closure position variation}
\entry{$h$}{Depth of grooves}
\entry{$p_{ref}$}{Inlet pressure}
\entry{$p_{vap}$}{Saturation pressure of water}
\entry{$v_{ref}$}{Inlet discharge velocity}
\entry{$v_{throat}$}{Velocity measured at the Venturi throat}
\entry{$v_{y}$}{Longitudinal component of flow velocity}
\entry{$v'_{y}$}{Flow velocity component parallel to the bottom wall Venturi slope}
\entry{$v'_{y}*$}{Undimensional flow velocity component parallel to the bottom wall Venturi slope}
\entry{$v'_{\infty}$}{Flow velocity component parallel to the bottom wall Venturi slope measured far from the plate wall}
\entry{$x$,$y$,$z$}{Cartesian coordinate system}
\entry{$x$,$y'$,$z'$}{Cartesian coordinate system inclined, parallel to the bottom wall Venturi slope}
\entry{$x^{*}$,$y^{*}$,$z^{*}$}{Dimensionless space variables}
\entry{$y'^{*}$,$z'^{*}$}{Dimensionless space variables}
\entry{$\delta_{v}$}{Viscous sublayer thickness}
\entry{$\lambda$}{Distance between the middle of two adjacent grooves}
\entry{$\nu$}{Water viscosity}
\entry{$\rho$}{Water density}
\entry{$\sigma$}{Cavitation number}
\entry{$^{*}$}{Symbol for undimensional variables: lengths are divided by $H_{throat}$ and velocities by $v'_{\infty}$}
\end{nomenclature}

\section{Introduction}

Cavitation is a crucial phenomenon encountered in fluid mechanics and for instance in turbomachinery domain. The unsteady character of the sheet cavity behavior on suction side of hydrofoils, on converging-diverging obstacles or on blades in turbines and propellers is responsible for many issues like erosion, noise and vibrations. However, in many industrial devices, cavitation inception can not be avoided, so the challenge consists in a better understanding of the two-phase flow dynamics once cavitation has appeared, in order to reduce its negative effects. Simple geometries like 2D foil sections, or converging-diverging steps like Venturi-type sections are used to approach sheet cavitation dynamics \cite{Barre2009}. A low pressure zone appears in the flow, downstream this type of obstacles. When this pressure becomes lower than the fluid saturation pressure, cavitation is induced and and a two-phase flow is obtained.

Sheet cavity dynamics can be described with a well-known cycle \cite{Knapp1955, Avellan1991, Stutz1997, Callenaere2001, Dular2004, Coutier2006}. Sheet cavity grows from the Venturi throat until a re-entrant jet appears in the sheet cavity closure, and flows upstream, near the wall, below the cavity. The re-entrant jet separates then the sheet cavity when it reaches the liquid-gas interface of the flow. A large vapour cloud is shed and convected downstream the flow, while the sheet cavity length is substantially reduced. Then, the cloud of vapor collapses in a higher pressure region and sustains the re-entrant jet due to cloud implosion, which later produces a new shedding of vapor cloud \cite{DeLange1998, Callenaere2001, Bergerat2012}. This cycle is characterized by the shedding frequency $\mathit{f}$. Such oscillatory behaviour of the sheet cavity is a witness of two-dimensional and three-dimensional cavity instabilities. These instabilities induce vibrations which are a disability for many industrial applications, as for inducers \cite{Bakir2003, Mejri2006, Campos2010}. 

Many studies have analyzed this sheet cavity cycle observed on cavitating hydrofoils or Venturi profiles. They have resulted in the definition of two types of oscillatory behaviors \cite{Callenaere2001, Sayyaadi2010}: the first one called cloud cavitation regime appears when the sheet cavity is long enough and generates periodical shedding as it has been described previously. But when the sheet cavity length is shorter because of a higher pressure level \cite{Callenaere2001} and/or a Reynolds number smaller than a critical value \cite{Keil2011, Keil2012}, no cloud cavitation shedding occurs. In this regime, the sheet cavity pulses but the cavity closure is always at the same location and no large scale detached cavitation structure is observed. In order to characterize these two regimes, a Strouhal number $\mathit{St_{L}}$ is defined (as it is presented in section~\ref{TwoCavitationRegimes}). For a sheet cavity regime, $\mathit{St_{L}\simeq}\,0.1$ while a cloud cavitation regime is characterized by $0.2\leq \mathit{St_{L}} \leq 0.4$. The cloud cavitation regime is more aggressive and leads to increased damages of the solid structures.. It is thus interesting to find a way to control the sheet cavity instability in order to limit erosion and/or noise. A passive control method of the sheet cavity regime has therefore to be evaluated by its ability to reduce the sheet cavity length, in order to favorise nearly stable regimes in front of periodical unsteady behaviors.      

Some studies have already been conducted to delay cavitation inception by modifying the foil surface roughness using local protuberances with different geometries or distributed irregularities \cite{Arndt1968, Arndt1981}. This modification acts on the turbulent boundary layer which drives the onset of sheet cavitation. However, in many industrial applications, the effect of developped cavitation on surface drag and/or performance is also of primary importance. Therefore, attention should be paid not only on delaying cavitation inception, but also on moderating fully developped cavitation. Only a few studies exist about the role of the surface condition in a fully developed cavitating flow. Some of these works show that roughness is able to decrease sheet cavity length, to increase oscillation frequency of the cavity or even to change cavitation regime \cite{Kawanami1997, Stutz2003b, Coutier2005}. The roughness distribution and geometry is a major characteristic of this passive method of cavitation control: tranversal or longitudinal grooves, with smooth or straight edges, for different depths or interval lengths \cite{Yongjian2010}.

The present study investigates the effect of different grooved suction side surfaces of a Venturi-type section with convergent and divergent angles respectively $18^{\circ}$ and $8^{\circ}$. Different grooved plates have been studied in order to emphasize the crucial geometric parameter of this organized roughness to obtain a passive control of cavitation. High speed visualizations of the cavitating flow show the effect of flow modifications near the wall on the sheet cavity development, and next, velocity measurements in non-cavitating conditions clarify the effect of the grooved surface on the flow structure.

\section{Experimental set-up}\label{ExperimentalSetup}

\begin{figure}
\centerline{\psfig{figure=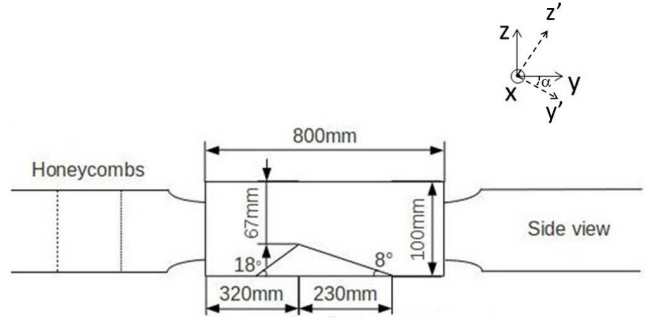,width=3.34in}}
  \caption{Test section of the experimental device using a Venturi profile (the origin of the cartesian coordinate system is located at the throat, in the middle of the test section wide, $\mathit{\alpha=}8^{\circ}$ is the angle for the LDV measurements system $\mathit{xy'z'}$ and $\mathit{H_{throat}=}67\,\mathrm{mm}$ is the reference length).}
\label{fig:TestSection}
\end{figure}

\begin{figure}
\centerline{\psfig{figure=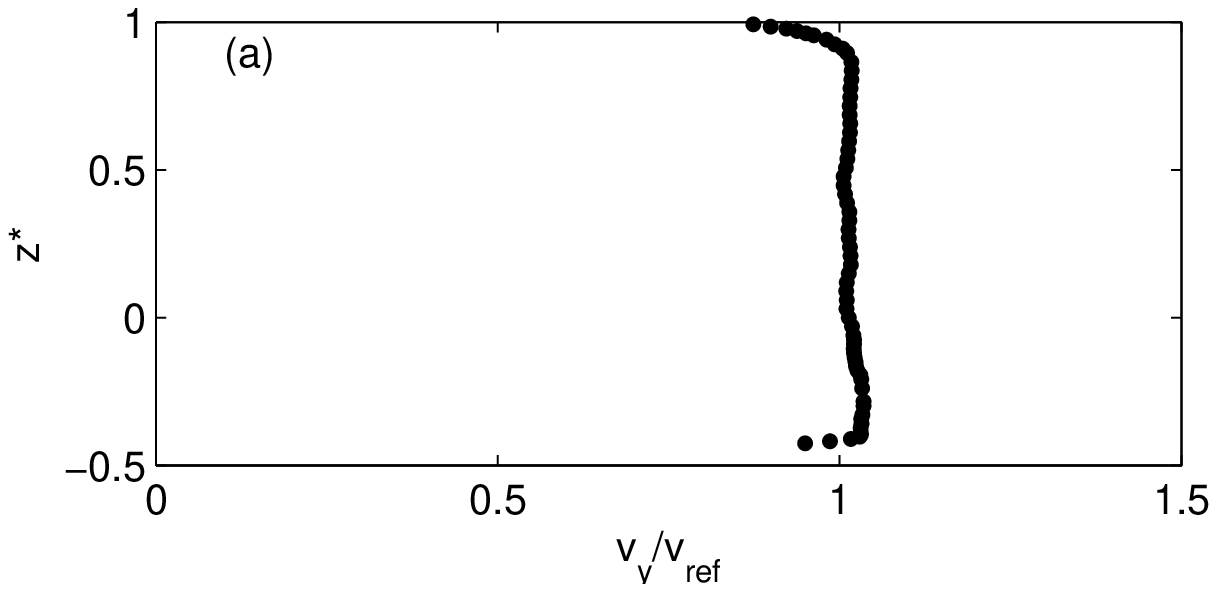,width=3.34in}}
\centerline{\psfig{figure=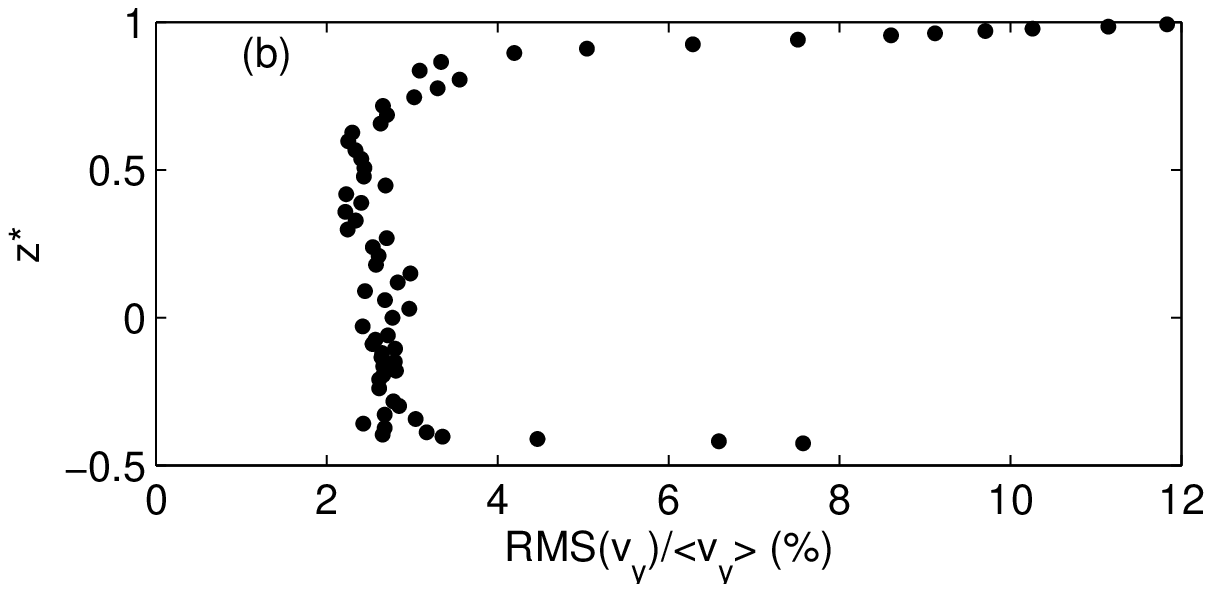,width=3.34in}}
  \caption{Characterisation of the flow in the test section inlet: (\textit{a}) undimensional longitudinal velocity profile $\mathit{v_{y}^{*}=v_{y}/v_{ref}}$ (the line represents the mean value of the velocity $\mathit{v_{ref}=}5.56\,\mathrm{m.s}^{-1}$) and (\textit{b}) the profile of the turbulence intensity $\mathit{\frac{RMS(v_{y})}{<v_{y}>}}$, where $<\mathit{v_{y}}>$ is the time-averaged velocity, $\mathit{y^{*}\simeq} -4.5$ and $\mathit{x^{*}=} 0$ (Results come from LDV measurements, presented in the section~\ref{LDV}).}
\label{fig:VitesseAmont}
\end{figure}

Experiments were conducted in the closed loop test rig of the DynFluid laboratory water tunnel. This rig is composed of two storage tanks with a capacity of $4\,\mathrm{m}^{3}$ each. The centrifugal pump can reach $1450\,\mathrm{rpm}$. The test-section for experiments is presented in Fig.~\ref{fig:TestSection}. The study flow volume, which is presented in Fig.~\ref{fig:TestSection}, measures $120\times 100\times 800\,\mathrm{mm}^{3}$. The bottom wall is made of a convergent ($18^{\circ}$ angle) and a divergent ($8^{\circ}$ angle), which results in a minimum height $\mathit{H_{throat}=}\,67\,\mathrm{mm}$ at the Venturi throat. The flow rate is fixed for all measurements at $\mathit{Q=}240\,\mathrm{m}^{3}.\mathrm{h}^{-1}$, which gives a maximun velocity at the Venturi throat $\mathit{v_{throat}}\simeq 8\,\mathrm{m.s}^{-1}$. Considering that the water viscosity in the operating conditions is $\mathit{\nu=}1.007~10^{-6}\,\mathrm{m}^{2}.\mathrm{s}^{-1}$, then the Reynolds number is $Re_{throat}\mathit{=v_{throat}H_{throat}/\nu}\simeq 5.5~10^{5}$, for the water at a temperature $\mathit{T=}\left(19\pm 1\right)^{\circ}\mathrm{C}$. Several honeycombs followed by a circular contraction provide the flow a velocity top hat profile upstream the Venturi with an inlet velocity equal to $\mathit{v_{ref}=}5.56\,\mathrm{m.s}^{-1}$, for all experimental configurations, with less than $3\%$ of turbulence intensity (Fig.~\ref{fig:VitesseAmont}), measured at a distance $\mathit{y=}300\,\mathrm{mm}$ upstream the Venturi throat (at an undimensional distance $\mathit{y^{*}=y/H_{throat}\simeq} -4.5$ from the throat). A vacuum pump can decrease the pressure in the test rig by decreasing pressure at the free surface in each storage tanks in order to acts on the cavitation number $\mathit{\sigma=(p_{ref}-p_{vap})/}(\frac{1}{2}\mathit{\rho v_{ref}}^{2})$, with $\mathit{p_{ref}}$ the pressure measured at $\mathit{y=}250\,\mathrm{mm}$ upstream the Venturi throat (at $\mathit{y^{*}=y/H_{throat}\simeq} -3.73$). The test section includes four plexiglas windows to permit three-dimensional visualisations of the flow. 

\begin{figure*}
\centerline{\psfig{figure=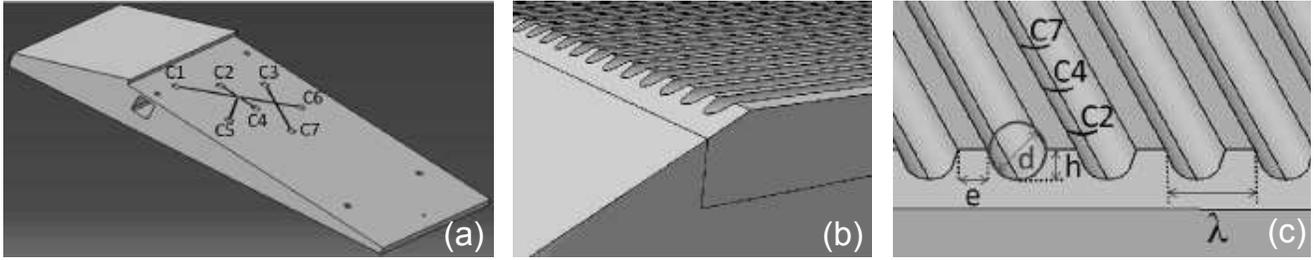,width=6.85in}}
    \caption{Venturi components: (\textit{a}) Venturi basis with $C1$ to $C7$ pressure sensors on the basis surface, (\textit{b}) junction between the Venturi basis and a plate and (\textit{c}) zoom of a grooved plate with the definition of the grooves geometric parameters ($\mathit{d}$ and $\mathit{h}$ are respectively the grooves diameter and depth, $\mathit{e}$ is the width of the ridge and $\mathit{\lambda}$ is the grooves wavelength).}
\label{fig:Plaques}
\end{figure*}

The bottom wall of the test section, in the divergent part of the Venturi, is made of interchangeable plates of dimensions $7\,\mathrm{mm}$ thick, $242\,\mathrm{mm}$ long, and $120\,\mathrm{mm}$ wide, to cover the entire Venturi surface downstream the Venturi throat. The junction between the plate and the Venturi basis is done upstream the Venturi throat to minimize its disturbances on the flow. So disturbances are expected to be smoothed before the flow comes on the throat, thanks to the favorable pressure gradient. Seven 2 MI-PAA KELLER absolute pressure sensors even out with the Venturi basis wall and distributed on the Venturi basis surface (Fig.\ref{fig:Plaques} and table~\ref{tab:PressureSensors}). These sensors, which are $4.5\,\mathrm{mm}$ in diameter, acquire pressure measurements during $1\,\mathrm{min}$ at a sampling rate of $1000\,\mathrm{Hz}$. Their sensitivity is $35\,\mathrm{mV.bar^{-1}}$. The plate is screwed on the basis by ensuring the sealing between both components with a vacuum grease. Table~\ref{tab:plates} presents the different grooved plates used for experiments. Plate $0$ will be considered as the reference case, with a smooth surface on the suction side of the Venturi. The other plates, from $1$ to $8$, have grooved surfaces. These grooves are made with a $1\,\mathrm{mm}$ or $2\,\mathrm{mm}$ diameter bur $\mathit{d}$, with a $\mathit{h}$ depth and the width of the ridge is $\mathit{e}$. The distance between the middle of two adjacent grooves is equal to $\mathit{\lambda=e+}2\sqrt{\mathit{dh-h}^{2}}$ for plates $1$ to $6$ and $\mathit{\lambda=e+d}$ for plates $7$ and $8$. Grooves open to the throat, so roughness effects already from the onset of the sheet cavity. All positions in the test section are expressed with undimensional values $x^{*}=x/H_{throat}$, $y^{*}=y/H_{throat}$ and $z^{*}=z/H_{throat}$, and the origin of the cartesian coordinate system is located in the throat, in the middle of the test section width.    

\begin{table}
  \caption{Characteristics of the studied plates used on the Venturi suction side ($\mathit{N}$ is the number of grooves) and description os symbols used in following graphics}
  \begin{center}
\def~{\hphantom{0}}
  \begin{tabular}{ccccccc}
  \hline
Plate & $\mathit{d}$ & $\mathit{h}$ & $\mathit{e}$ & $\mathit{\lambda}$ & $\mathit{N}$ & Symbol\\[3pt]
 & ($\mathrm{mm}$) & ($\mathrm{mm}$) & ($\mathrm{mm}$) & ($\mathrm{mm}$) & & \\
 \hline
0 & 0 & 0 & 0 & 0~~~ & 0 & \textcolor{black}{$\times$}\\
1 & 1 & ~~~0.15 & ~~0.1 & 0.81 & 147 & \textcolor{black}{$\bigstar$}\\
2 & 1 & ~~~0.25 & ~~0.1 & 0.97 & 124 & \textcolor{black}{$\triangle$}\\
3 & 1 & ~~0.5 & ~~0.1 & 1.1~ & 109 & \textcolor{black}{$\blacktriangle$}\\
4 & 1 & ~~0.5 & 1 & 2~~~ & 60 & \textcolor{black}{$\bullet$}\\
5 & 2 & ~~0.5 & ~~0.1 & 1.83 & 65 & \textcolor{black}{$\bigvarstar$}\\
6 & 2 & 1 & ~~0.1 & 2.1~ & 57 & \textcolor{black}{$+$}\\
7 & 2 & 2 & 1 & 3~~~ & 40 & \textcolor{black}{$\diamondsuit$}\\
8 & 2 & 2 & ~~0.1 & 2.1~ & 57 & \textcolor{black}{$\Box$}\\  
\hline
  \end{tabular}
  \label{tab:plates}
  \end{center}
\end{table}

\begin{table}
  \caption{Pressure sensors on the Venturi suction side}
  \begin{center}
\def~{\hphantom{0}}
  \begin{tabular}{ccccc}
  \hline
Pressure sensor & $\mathit{y^{'}}(\mathrm{mm})$ & $\mathit{y^{'*}}$ & $\mathit{x}(\mathrm{mm})$ & $\mathit{x^{*}}$ \\[3pt]
\hline
C1 & ~16.75 & 0.25 & ~30  & ~0.4478\\
C2 & ~33.50 & 0.5~ & ~~0 & ~0~~~~~\\
C3 & ~50.25 & 0.75 & -30 & -0.4478\\
C4 & ~67~~~ & 1~~~ & ~~0 & ~0~~~~~\\
C5 & ~67~~~ & 1~~~ & ~30 & ~0.4478\\
C6 & ~83.75 & 1.25 & -30 & -0.4478\\
C7 & 100.5 & 1.5~ & ~~0 & ~0~~~~~\\ 
\hline
  \end{tabular}
  \label{tab:PressureSensors}
  \end{center}
\end{table}

The test of these grooved plates and their comparison with the results obtained with the smooth reference plate $0$ allow to evaluate the effects the effect of the surface condition of the Venturi bottom wall on the flow dynamics near the wall, with or without cavitation.

\section{Effect of grooved surfaces on cavitating flow}\label{sec:CavitantFlow}

Visualisations are conducted for each plate (the smooth and the grooved ones), for different values of cavitation number $\mathit{\sigma}$ defined as $\mathit{\sigma=(p_{ref}-p_{vap})/}(\mathrm{\frac{1}{2}}\mathit{\rho v_{ref}}^{2})$. In this definition, $\mathit{p_{ref}}$ and $\mathit{v_{ref}}$ represent the inlet pressure and velocity of the test section, measured $250\,\mathrm{mm}$ upstream the Venturi throat, while $\mathit{p_{vap}}$ is the vapor pressure at temperature $\mathit{T=}292\,\mathrm{K}$ of the fluid with a density $\mathit{\rho}$. Temperature deviation during measurements is small enough (about $1\,\mathrm{K}$) to neglect temperature effect on results but in order to have precise values of $\mathit{\sigma}$, the flow temperature is measured at each acquisition (with an uncertainty of $0.1\,\mathrm{K}$). Temperature for all measurements is then considered as the ambient temperature and $\mathit{p_{vap}}\simeq 2200\,\mathrm{Pa}$. For each studied configuration, $\mathit{\sigma}$ ranges from $0.9$ to $1.7$ with $1\%$ of uncertainty.

\subsection{Sheet cavity visualisations}

Visualisations of the sheet cavity developed on the Venturi consist in the acquisition of $4000$ images for each configuration (for a given cavitation number and a given grooved plate geometry), in the $\left(\mathrm{y0z}\right)$ plane. A CamRecord $600$ Optronics camera is used with a Zeiss Makro-planar T* which has a focal length of $100\,\mathrm{mm}$. This high-speed CMOS camera records images with a $1280 \times 512$ pixels resolution, a rate of $1000\,\mathrm{fps}$ and an exposure time equal to $0.167\,\mathrm{\mu s}$. The pixels size is $12\,\mathrm{\mu m} \times 12\,\mathrm{\mu m}$ for an active area of $15.35\,\mathrm{mm} \times 12.29\,\mathrm{mm}$. The flow is illuminated by a Super Long Life Ultra Bright (SLLUB) White Led Backlight from Phlox on a $200\,\mathrm{mm} \times 200\,\mathrm{mm}$ light output area. Its minimal luminance is $3000\,\mathrm{cd.m}^{-2}$ in continuous mode and its uniformity is $99.54\%$. This light source gives then images integrated on the entire width of the test section, but the focusing is made as far as possible in the middle of the section, in order to prevent side walls effects.

\subsection{Effects of the Venturi surface condition on the sheet cavity size}\label{SheetCavitySize}

\begin{figure}
\centerline{\psfig{figure=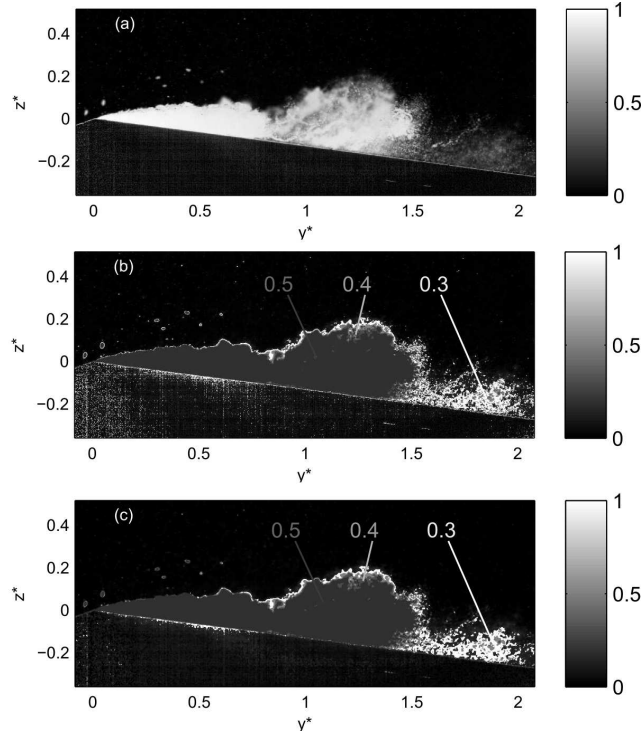,width=3.34in}}
  \caption{Images processing for the sheet cavity length measurements: (\textit{a}) normalized instantaneous image, (\textit{b}) image binarization with different threshold levels 0.5, 0.4 or 0.3 and (\textit{c}) median filter applied on the binarized image.}
\label{fig:Binarisation}
\end{figure}

\begin{figure*}
\centerline{\psfig{figure=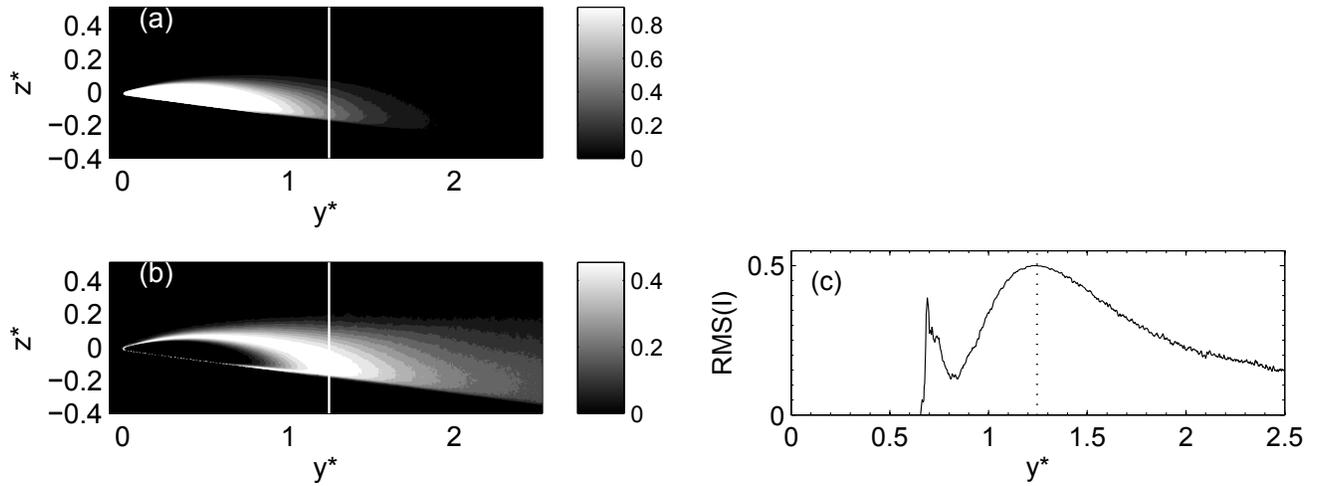,width=6.85in}}
  \caption{Sheet cavity on the smooth plate $0$ for $\mathit{\sigma=}1.18$: (\textit{a}) normalized average image, (\textit{b}) root mean square of normalized images (the line represents the undimensional sheet cavity length $\mathit{L^{*}=L/H_{throat}}\simeq 1.62$) and (\textit{c}) a profile of the root mean square of normalized images $\mathrm{RMS}(\mathit{I})$ plotted in z*=0.}
\label{fig:AverageRMS}
\end{figure*}

\begin{figure}
\centerline{\psfig{figure=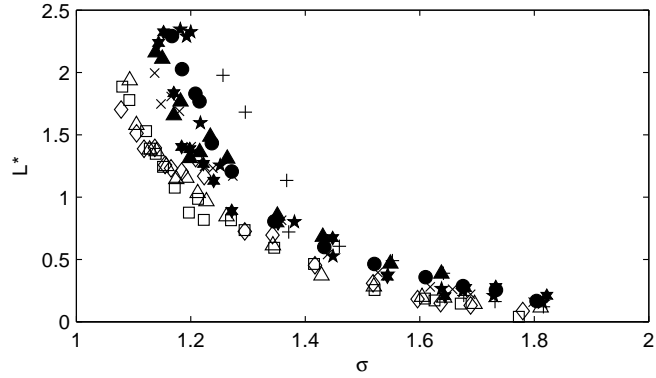,width=3.45in}}
  \caption{Undimensional sheet cavity length $\mathit{L^{*}=L/H_{throat}}$ according to the cavitation number $\mathit{\sigma}$: \textcolor{black}{$\times$}: plate 0, \textcolor{black}{$\bigstar$}: plate 1, \textcolor{black}{$\triangle$}: plate 2, \textcolor{black}{$\blacktriangle$}: plate 3 , \textcolor{black}{$ + $}: plate 4, \textcolor{black}{$\bigvarstar$}: plate 5, \textcolor{black}{$+$}: plate 6, \textcolor{black}{$\diamondsuit$}: plate 7, \textcolor{black}{$\Box$}: plate 8.}
\label{fig:Ladim_sigma}
\end{figure}

In order to determine the cavity mean length, mean values and standard deviation of the grey levels are calculated with the $4000$ images recorded in each flow configuration, in the $\left(\mathrm{y0z}\right)$ plane (side views of the attached sheet cavity dynamics). The instantaneous images are normalised by a reference image. Normalised images are then binarised with a threshold level equal to 0.4 and the noise is filtered with a median filter as presented in Fig.~\ref{fig:Binarisation}. The thresehold level is chosen with the Otsu's method~\cite{Otsu1979} and validated with comparisons with the initial normalised images (Fig.~\ref{fig:Binarisation}(b)). The mean cavity length extends from the Venturi throat to the cloud cavitation separation location. This closure of the sheet cavity is detected by localising the maximum value of the standard deviation of the grey levels according to \cite{Dular2004} (Fig.~\ref{fig:AverageRMS}). Figure~\ref{fig:Ladim_sigma} shows the evolution of the cavity mean length $L$ according to the cavitation number $\mathit{\sigma}$, for all plates. The uncertainty of $\mathit{L}$ is less than $2\%$ (by evaluating the effect of images processing) and $\mathit{\sigma}$ is calculated with $1\%$ of uncertainty. Plates $1$,$3$, $4$ and $5$ provide a sheet cavity length close to the one obtained with the reference plate $0$. Conversely, plates $2$, $6$, $7$ and $8$ lead to significant differences of the cavity lengths. The results obtained with these three plates are thus detailed hereafter. It can be observed that for the plate $2$, the cavity mean length decreases as it has been encountered in litterature, for random distributed roughness \cite{Coutier2005}. This plate has a large number of grooves ($\mathit{N=}124$ and $\mathit{\lambda =}0.97\,\mathrm{mm}$) and is characterized by a small depth $\mathit{h=}\,0.25\,\mathrm{mm}$. But other types of grooves have the opposite effect, i.e. an increase of the mean length of the sheet cavity, as we can see for the plate $6$, which has a small number of grooves ($\mathit{N=}57$ and $\mathit{\lambda=}2.1\,\mathrm{mm}$) with a large depth $\mathit{h=}\,1\,\mathrm{mm}$. This difference between the effects of surface condition is larger for small cavitation numbers. When $\mathit{\sigma}$ is greater than $1.2$ i.e. for small sheet cavities, the influence of the surface condition on the mean cavity length seems to be less important. The major difference between plates $2$ and $6$, is the depth parameter $\mathit{h}$, the diameter $\mathit{d}$ and the number of grooves on the bottom wall of the Venturi. In comparison with plates $3$, $4$ and $5$ which do not affect significantly the sheet cavity length with respect to the reference plate $0$, the depth $\mathit{h}$ seems to be the criterion that drives the change of effects on the sheet cavity. Indeed, the plate $2$ is one of the plates which has the smallest depth. Plate $1$ has an even smaller depth, but this plate seems to have no effect on the cavitation dynamics. The depth of plate $1$ has been chosen in order to be of the order of the viscous sublayer thickness $\mathit{\delta_{v}}$ calculated on a flat plate (zone of the boundary layer where the velocity increases linearly from the wall, beyond this zone, the velocity increases logarithmically with the height). In the case of a plane plate\cite{Schlichting2004}, $\mathit{\delta_{v}}$ is defined by:

\begin{equation}
\frac{\mathit{\delta_{v}}}{\mathit{y^{*}}}=\frac{50}{Re_{\mathit{y^{*}}}\sqrt{\frac{\mathit{C_{f}}}{2}}}
\end{equation} 

where $\mathit{Re_{y}}$ is the critical Reynolds number which determines the distance $\mathit{y^{*}}$  from the beginning of the plate where the boundary layer becomes turbulent ($Re_{\mathit{y^{*}}}=5~10^{5}$ for a plane plate) and $\mathit{C_{f}}$ is the coefficient of friction expressed as:

\begin{equation}
\mathit{C_{f}}=2\left(\frac{\mathrm{K}}{\log Re_{\mathit{y^{*}}}}\mathrm{G}\left(\log Re_{\mathit{y^{*}}}\right)\right)^{2} 
\end{equation}

with $\mathrm{K}=0.41$ the Karman constant and $\mathrm{G}$ a function that gives $\mathrm{G}\left(\log\left(Re_{\mathit{y^{*}}}\right)\right)=1.5$ in the studied zone $10^{5}<Re_{\mathit{y^{*}}}<10^{6}$ \cite{Schlichting2004}.

If we assume that our experimental set-up used is a fat plate, without pressure gradient, $\mathit{\delta_{v}}$ is inferior to $0.20\,\mathrm{mm}$ ($\mathit{\delta_{v}}/\mathit{H_{throat}}\simeq 0.003$) at the end of the plate (at $\mathit{y^{*}}\simeq 3.4$). This calculation allows us to obtain an order of magnitude of the viscous sublayer thickness of our experimental set-up, even if we have to consider the adverse pressure gradient and the effects of grooves on the boundary layer development to have a more accurate value of the viscous sublayer thickness.\\

Indeed, results obtained with plate $1$ (with a grooves depth inferior to the viscous sublayer thickness of a plane plate without pressure gradient $\mathit{\delta_v}$) are identical to the one measured with plate $0$. Conversely, if the depth of the grooves is too large, then the sheet cavity length increases. 

However, if the depth is increased up to $\mathit{h}=2\,\mathrm{mm}$, results are different: for plates $7$ and $8$, we can see a decrease of the sheet cavity length, like for the plate $2$ which presents a small depth. These two plates $7$ and $8$ are special because the geometry of the grooves is different. Grooves for plates $1$ to $6$ are arcuate hollows while grooves are cylindrical gutters for plates $7$ and $8$ ($\mathit{h}>\frac{\mathit{d}}{2}$). This geometry leads to a decrease of the sheet cavity length. This phenomenon may be related to the fluid flow inside the grooves, which may be different with plate $6$, compared with plates $7$ and $8$. Indeed, the number of grooves $\mathit{N}$ is the same for these three plates. It can thus be deduced that this parameter is not crucial for the effect of grooves on the sheet cavity length. Only the depth $\mathit{h}$ is different from plate $6$ to plates $7$ and $8$. It implies that a reduction or an increase of the sheet cavity length can both be obtained with appropriate grooves, depending on their depth $\mathit{h}$.\\

\begin{figure}

\centerline{\psfig{figure=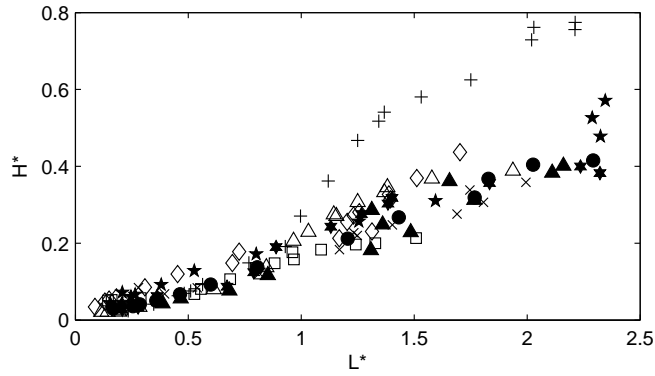,width=3.45in}}
  \caption{Undimensional sheet cavity mean height $\mathit{H^{*}=H/H_{throat}}$ according to the undimensional sheet cavity mean length $\mathit{L^{*}=L/H_{throat}}$: \textcolor{black}{$\times$}: plate 0, \textcolor{black}{$\bigstar$}: plate 1, \textcolor{black}{$\triangle$}: plate 2, \textcolor{black}{$\star$}: plate 3 , \textcolor{black}{$ + $}: plate 4, \textcolor{black}{$\bigvarstar$}: plate 5, \textcolor{black}{$+$}: plate 6, \textcolor{black}{$\diamondsuit$}: plate 7, \textcolor{black}{$\Box$}: plate 8.}
\label{fig:Ladim_Hadim}
\end{figure}

The former analysis was focused on the effects of the grooves on the sheet cavity length. In addition, the effects on the mean height of the sheet cavity are also investigated hereafter: for that purpose, the maximum height of the sheet cavity is detected. The calculation of the mean value of all these maximum heights (which can variate in position) gives information about the shape of the sheet cavity, according to the grooved surfaces. Figure~\ref{fig:Ladim_Hadim} displays the undimensional sheet cavity mean height $\mathit{H^{*}=H/H_{throat}}$ (with $2\%$ of uncertainty) according to the undimensional sheet cavity mean length $\mathit{L^{*}=L/H_{throat}}$, which depends on the cavitation number, for all different studied plates. It can be observed that results obtained with plate $6$ are significantly different from all other data. As grooved plates introduce a three-dimensional geometry with different grooves wavelengths, these Venturi surfaces induce a changing in the sheet cavity dynamics: the grooves wavelength forces the flow, and especially the inception of the sheet cavity, in the grooves hollow, but also the development of the sheet cavity or the re-entrant jet. This phenomen, with a three-dimensional wavelength imposed to the flow is different from the smooth case, with the plate $0$. The cavity closure line is actually inclined \cite{DeLange1996} and the cloud cavitation shedding is linked to three-dimensional instabilities \cite{Laberteaux1998, Duttweiler1998, Foeth2008}, three-dimensional effects are then to be taken into account to evaluate the efficiency of the passive control. As each plate presents a different spanwise distribution, we can say that depth is a crucial parameter but that the grooves wavelength $\mathit{\lambda}$ is also important. Plates $6$, $7$ and $8$ have largest wavelength ($\mathit{\lambda}>2\,\mathrm{mm}$ and $\mathit{N}<60$). For small sheet cavities characterised by $0< \mathit{L^{*}} \leq 1$, which corresponds to $1.2\leq \mathit{\sigma} \leq 1.9$, all curves are nearly superimposed, which suggests that grooves play a less crucial role.

\begin{figure}
\centerline{\psfig{figure=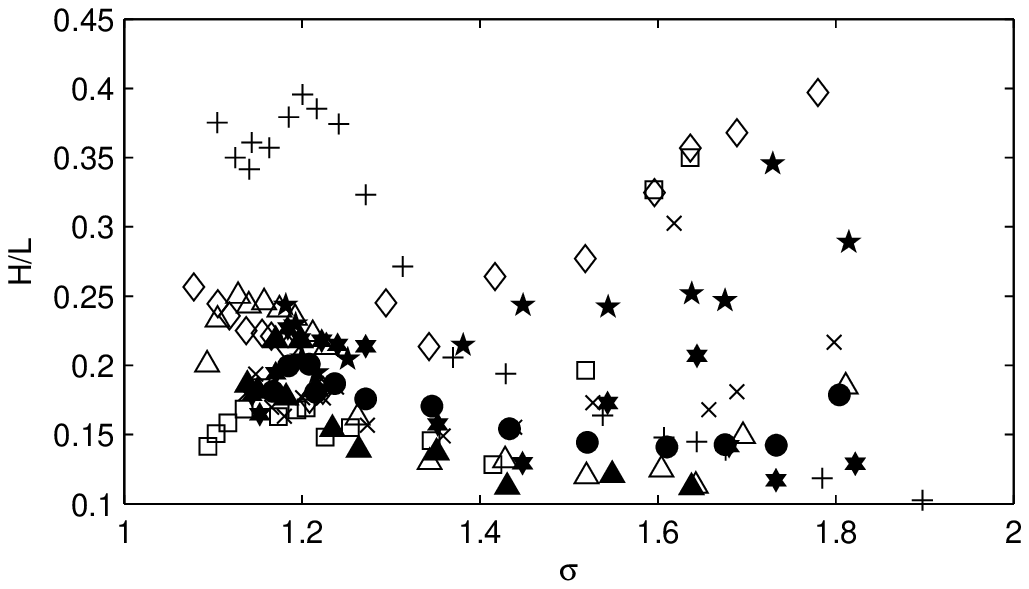,width=3.45in}}
  \caption{Aspect ratio $\mathit{H/L}$ according to the cavitation number $\mathit{\sigma}$: \textcolor{black}{$\times$}: plate 0, \textcolor{black}{$\bigstar$}: plate 1, \textcolor{black}{$\triangle$}: plate 2, \textcolor{black}{$\star$}: plate 3 , \textcolor{black}{$ + $}: plate 4, \textcolor{black}{$\bigvarstar$}: plate 5, \textcolor{black}{$+$}: plate 6, \textcolor{black}{$star$}: plate 7, \textcolor{black}{$\Box$}: plate 8.}
\label{fig:AspectRatio}
\end{figure}

Figure~\ref{fig:AspectRatio} which presents the aspect ratio $\mathit{H/L}$  according to the cavitation number for all grooved plates, shows that most of grooved plates present large variations of the aspect ratio values with different cavitation numbers while the reference plate case $0$ leads to a nearly constant aspect ratio. The grooved plates $3$, $4$ and $5$ for which the sheet cavity length is almost equal to the reference case, for all cavitation numbers, are characterized by an aspect ratio $0.1\leq \frac{\mathit{H}}{\mathit{L}} \leq 0.2$. These values are still close to the reference case. On the other hand, the plate $1$, with a grooves depth $h$ too small to modify the sheet cavity length, has an aspect ratio very different from the smooth plate $0$. This result shows that the grooves have effects not only on the sheet cavity length but also on its shape. Plates $7$ and $8$ present an evolution of the aspect ratio $\mathit{H/L}$ similar to the plate $1$. The most obvious observation is that the evolution is inversed for plate $6$. Indeed, for plate $6$, the ratio $\mathit{H/L}$ is high for $1\leq \mathit{\sigma} \leq 1.4$ and small for $1.4\leq \mathit{\sigma} \leq 1.9$ ($0.2\leq \frac{\mathit{H}}{\mathit{L}} \leq 0.4$ for the first range of $\mathit{\sigma}$ and $0.1\leq \frac{\mathit{H}}{\mathit{L}} \leq 0.2$ for the second), while, for plates $7$ and $8$, $\mathit{H/L}$ is first in the range of $0.1\leq \frac{\mathit{H}}{\mathit{L}} \leq 0.2$ when $1\leq \mathit{\sigma} \leq 1.4$ and then $0.2\leq \frac{\mathit{H}}{\mathit{L}} \leq 0.4$ when $1.4\leq \mathit{\sigma} \leq 1.9$. Cavitation dynamics is thus completely different for plates $6$ and other plates, which have a smaller grooves wavelength $\mathit{\lambda}$, excluding plates $7$ and $8$ for which the grooves geometry is quite different. It confirms that the crucial parameter in the action of the grooves is $\mathit{i)}$ the depth, which may modify the re-entrant jet progression, $\mathit{ii)}$ the wavelength, which can force the cavitation dynamics in the spanwise component. As it was reminded in the introduction, the cavity regime depends on the mean size of the sheet cavity \cite{Callenaere2001}. An analysis of the Strouhal number based on the characteristic frequency of the sheet cavity is conducted to determine the effects of grooved surfaces on the unsteady behavior of the cavity.

\subsection{Pressure measurements}

In order to analyze the different cavitation dynamics detected with the grooved surfaces, pressure measurements have been acquired at the bottom wall of the Venturi. Relative pressure measurements have been first conducted with no velocity and no use of the vacuum pump to define the zero setting state for each sensor.

\begin{figure*}
\centerline{\psfig{figure=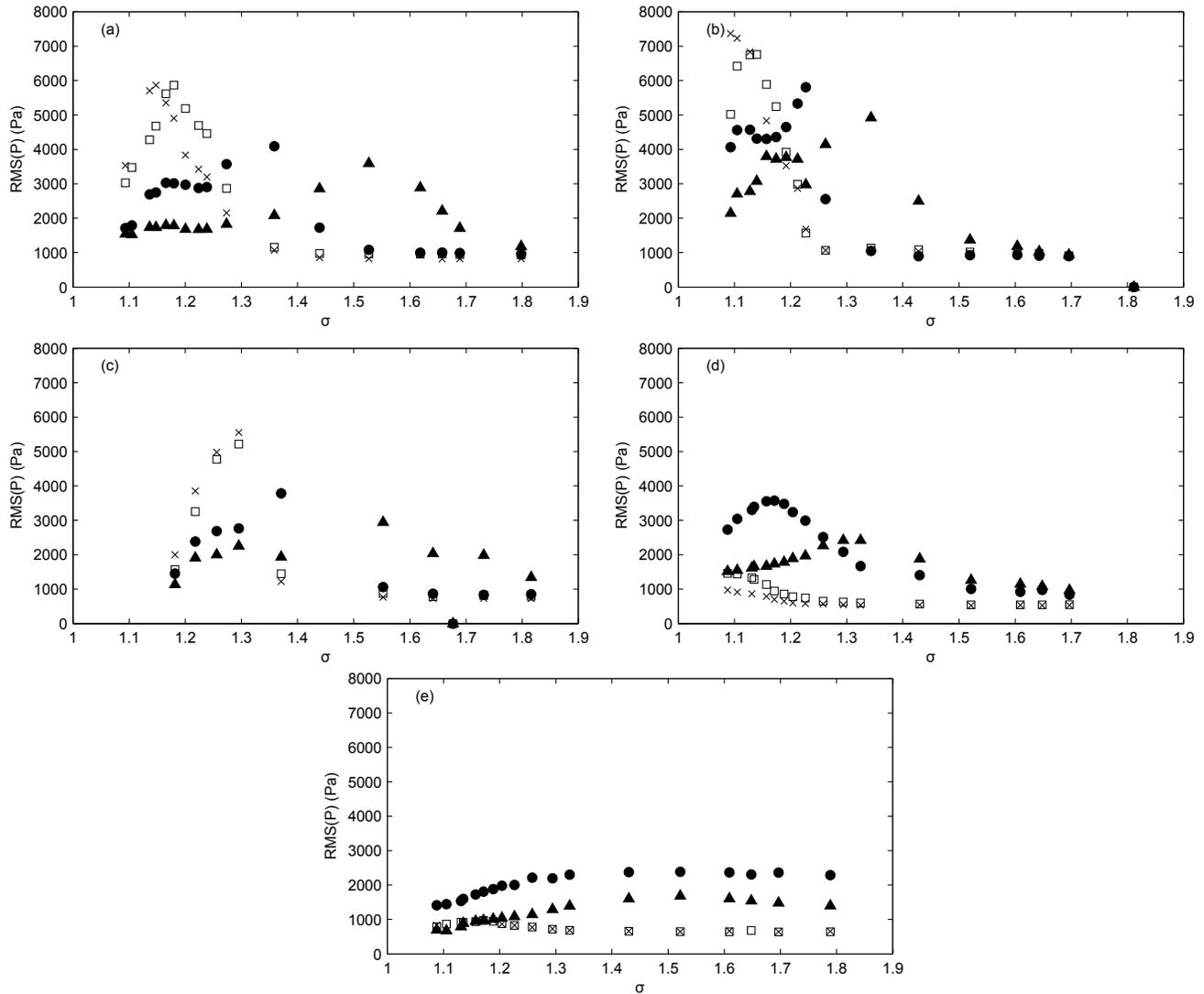,width=6.85in}}
  \caption{Pressure fluctuations $\mathrm{RMS}(\mathit{P})$ on the Venturi bottom wall according to the cavitation number $\mathit{\sigma}$, at different distances from the ventrui throat:($a$) for plate $0$,($b$) for plate $2$, ($c$) for plate $6$, ($d$) for plate $7$ and ($e$) for plate $8$. \textcolor{black}{$\blacktriangle$}: C1, \textcolor{black}{$\bullet$}: C2, \textcolor{black}{$\Box$}: C4, \textcolor{black}{$\times$}: C6. (see positions of sensors in Table~\ref{tab:PressureSensors})}
\label{fig:RMSPressureCavitant_sigma}
\end{figure*}

The analysis of the pressure fluctuations obtained with plates $0$, $2$, $6$, $7$ and $8$, on the Fig.~\ref{fig:RMSPressureCavitant_sigma}, indicates that the maximum amplitude of fluctuations of each pressure sensor corresponds to the sheet cavity closure \cite{Callenaere2001}. For example, for $\mathit{\sigma=}1.5$, the undimensional sheet cavity length is $\mathit{L^{*}}<0.5$ on the plate $0$ (Fig.~\ref{fig:Ladim_sigma}). The sensor $C1$ is the only sensor located in the sheet cavity. Fluctuations of pressure measurements is thus an indicator of sheet cavity length. This figure confirms the results obtained with the image processing: if we look at the sensor $C1$, the maximum of pressure fluctuations is at $\mathit{\sigma=}1.5$ for the smooth plate $0$, $\mathit{\sigma}<1.5$ for the plate $2$ and $\mathit{\sigma}>1.5$ for the plate $6$, with larger grooves. This result shows that the length of sheet cavities is smaller on plate $2$ than on plate $0$ and a longer sheet cavity is obtained on plate $6$ than on the others. As it was observed for the plate $8$ previously, pressure fluctuations are a witness of the cavity regime changing. Indeed, in the Fig.~\ref{fig:RMSPressureCavitant_sigma}(e), which represents results for the plate $8$, no clear maximum of $\frac{\mathrm{RMS}}{<\mathit{P}>}$ can be identified, so we are not able to detect the sheet cavity length by analyzing pressure measurements. This is an other indicator of the sheet cavity regime. The difference between a sheet cavity regime and a cloud cavitation regime can thus be demonstrated also by the analysis of the pressure measurements.

\subsection{Two cavitation regimes}\label{TwoCavitationRegimes}

\begin{figure}
\centerline{\psfig{figure=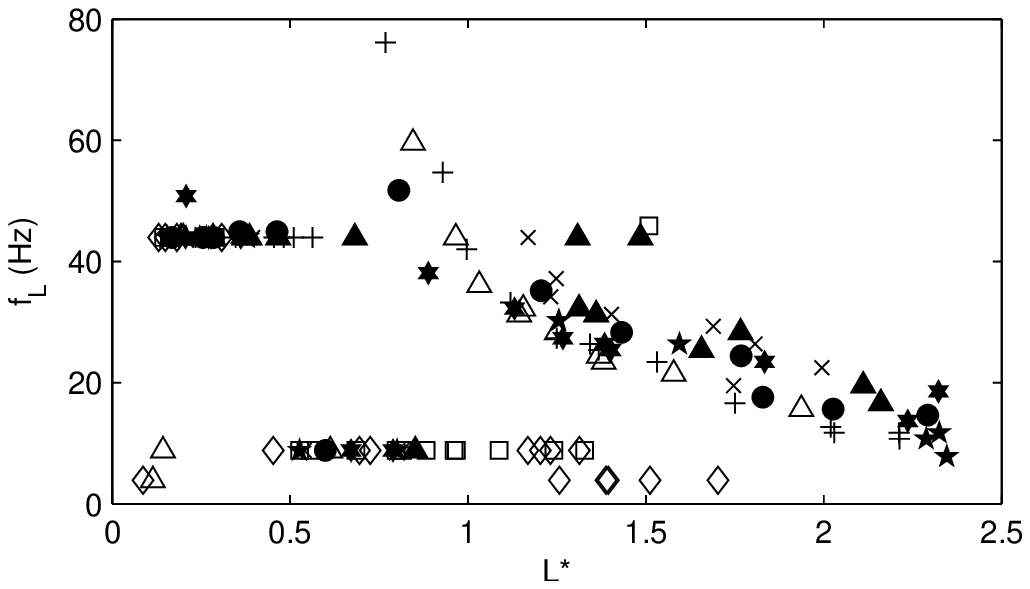,width=3.45in}}
  \caption{Frequency of the sheet cavity length oscillation $\mathit{f_{L}}$ according to the undimensional sheet cavity mean length $\mathit{L^{*}}$: \textcolor{black}{$\times$}: plate 0, \textcolor{black}{$\bigstar$}: plate 1, \textcolor{black}{$\triangle$}: plate 2, \textcolor{black}{$\star$}: plate 3 , \textcolor{black}{$ + $}: plate 4, \textcolor{black}{$\bigvarstar$}: plate 5, \textcolor{black}{$+$}: plate 6, \textcolor{black}{$\diamondsuit$}: plate 7, \textcolor{black}{$\Box$}: plate 8.}
\label{fig:fL_L}
\end{figure}

The fluctuations of the two-phase area in the sheet cavity regime or the shedding of the cloud of vapor in the cloud cavitation regime are both periodical. The frequency of these phenomena is determined by studying the evolution of the grey levels at the sheet cavity closure position (by selecting the vertical line of pixels situated in the mean position of the sheet cavity closure determined previously). The power spectral density of this sheet cavity length evolution allows to determine the characteristic frequency of the sheet cavity according to the mean cavity length for each studied case of Venturi profile grooved surface (Fig.\ref{fig:fL_L}), with $0.2\%$ of uncertainty.

\begin{figure*}
\centerline{\psfig{figure=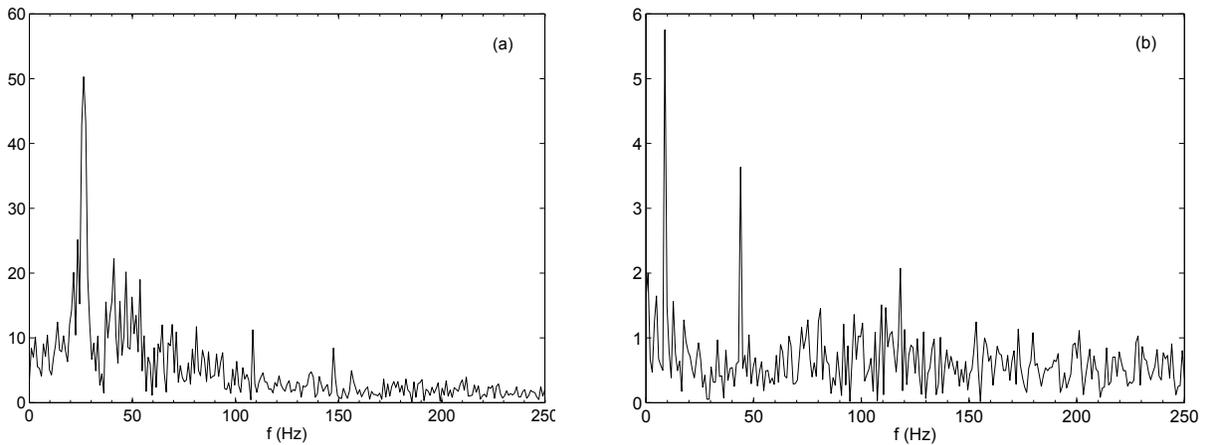,width=6.85in}}
  \caption{Frequency spectrum of the variation of the gray level in the closure of the sheet cavity, on the plate $0$: (a) for $\mathit{\sigma=}1.17$, (b) for $\mathit{\sigma=}1.44$}
\label{fig:SpectresLisse}
\end{figure*}

\begin{figure*}
\centerline{\psfig{figure=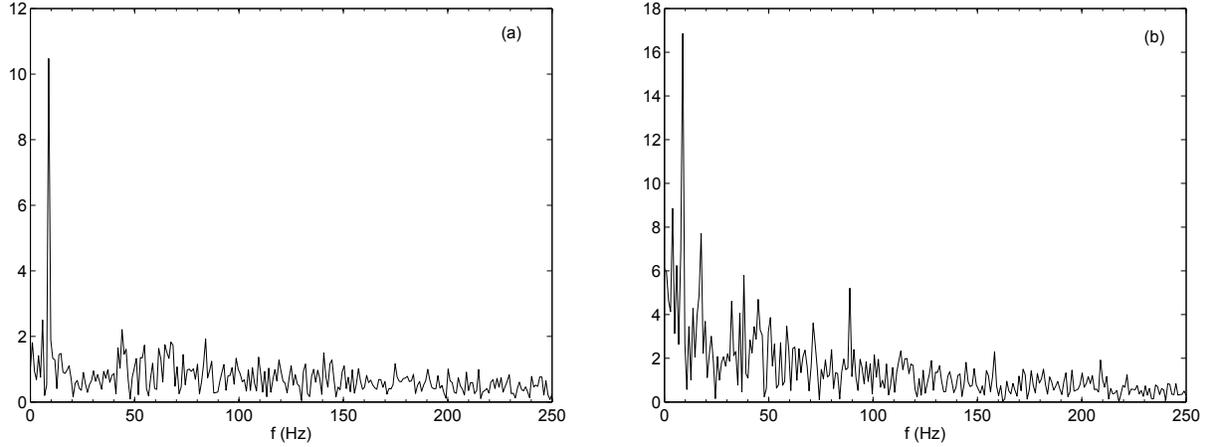,width=6.85in}}
  \caption{Frequency spectrum of the variation of the gray level in the closure of the sheet cavity, for $\mathit{\sigma=}1.17$ for: (a) the plate $7$, (b) the plate $8$}
\label{fig:Spectresrug7et8}
\end{figure*}

Here, the frequency difference between the smooth Venturi surface and the other grooved plates is greater when the mean cavity length is large (it means for small cavitation numbers). It can be observed in the range $0<\mathit{L^{*}}<1$, which corresponds to $1.3<\mathit{\sigma}<1.9$, two peaks on the frequency spectrum, with similar amplitudes, at $43.95\,\mathrm{Hz}$ and $8.79\,\mathrm{Hz}$. These two frequencies are harmonics and are both present in this studied range, as it is shown in Fig.~\ref{fig:SpectresLisse} for the plate $0$. This frequency of $43.95\,\mathrm{Hz}$ is the blade passing frequency of the centrifugal pump used to impose the flow rate: as the pump with $5$ blades operates at $528\,\mathrm{rpm}$, the blade passing frequency is $44\,\mathrm{Hz}$. It means that smallest cavities oscillate then at the blade passing frequency. For largest sheet cavities, so for a cavitation number $\mathit{\sigma}$ in the range $1<\mathit{\sigma}<1.3$, there is only one peak on the frequencies spectrum (Fig.\ref{fig:SpectresLisse} (a) presents results for $\mathit{\sigma=}1.17$). This peak corresponds to the shedding frequency of the cloud cavitation and it decreases when the sheet cavity length increases. Plates $7$ and $8$ lead to a different feature for the frequency derived from the grey level analysis in the closure of the sheet cavity. The cavitation number $\mathit{\sigma}$ has no effect on the frequency peak of spectra, which is always $\mathit{f_{L}=}43.95\,\mathrm{Hz}$. This result shows again that the cavitation dynamics is particular for these two plates $7$ and $8$. Figure~\ref{fig:Spectresrug7et8} shows frequency spectra obtained for plate $7$ (Fig.~\ref{fig:Spectresrug7et8}(a)) and plate $8$ (Fig.~\ref{fig:Spectresrug7et8}(b)) for $\mathit{\sigma=}1.17$.  
 
\begin{figure}
\centerline{\psfig{figure=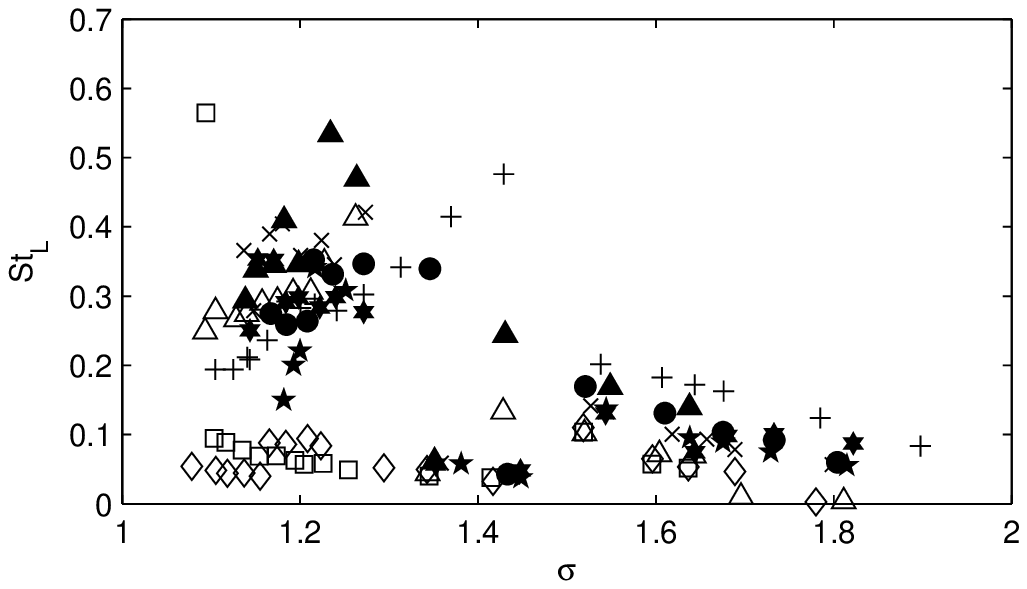,width=3.45in}}
  \caption{Strouhal number $\mathit{St_{L}}=\frac{\mathit{Lf_{L}}}{\mathit{v_{throat}}}$ according to the cavitation number $\mathit{\sigma}$: \textcolor{black}{$\times$}: plate 0, \textcolor{black}{$\bigstar$}: plate 1, \textcolor{black}{$\triangle$}: plate 2, \textcolor{black}{$\star$}: plate 3, \textcolor{black}{$ + $}: plate 4, \textcolor{black}{$\bigvarstar$}: plate 5, \textcolor{black}{$+$}: plate 6, \textcolor{black}{$\diamondsuit$}: plate 7, \textcolor{black}{$\Box$}: plate 8.}
\label{fig:St_sigma}
\end{figure}

The characteristic frequency of the sheet cavity is also represented on the Fig.~\ref{fig:St_sigma} with the undimensional Strouhal number $\mathit{St_{L}=Lf_{L}/v_{throat}}$, with $6\%$ of uncertainty. If we look at all plates excluding plates $7$ and $8$, we can observe that for $1\leq \mathit{\sigma} \leq 1.3$, the Strouhal number is around $0.3$, as it has been reported in previous experiments in \cite{Coutier2006}. This range of cavitation numbers leads to a cloud cavitation regime, characterized by large fluctuations of the cavity closure position and large cloud cavitation shedding. On the other hand, for $1.3\leq \mathit{\sigma} \leq 1.9$, the Strouhal number decreases until reaching a value smaller than $0.15$. This range corresponds to a sheet cavity regime, for which there is no convected cloud cavitation, the cavity closure position is almost at a constant distance from the Venturi throat. In the sheet cavity regime, the sheet cavity pulses. These two types of variation of the Strouhal number exist for grooved plates but also for the smooth one (plate $0$), and have been brought out by other authors \cite{Farhat1994, Dular2012}. It can be also observed in Fig.~\ref{fig:St_sigma} that the grooved plates surfaces can modify the Strouhal number. Indeed, smooth surface plate $0$ presents Strouhal numbers larger than grooved plates, essentially for small cavitation numbers.

\begin{figure*}
\centerline{\psfig{figure=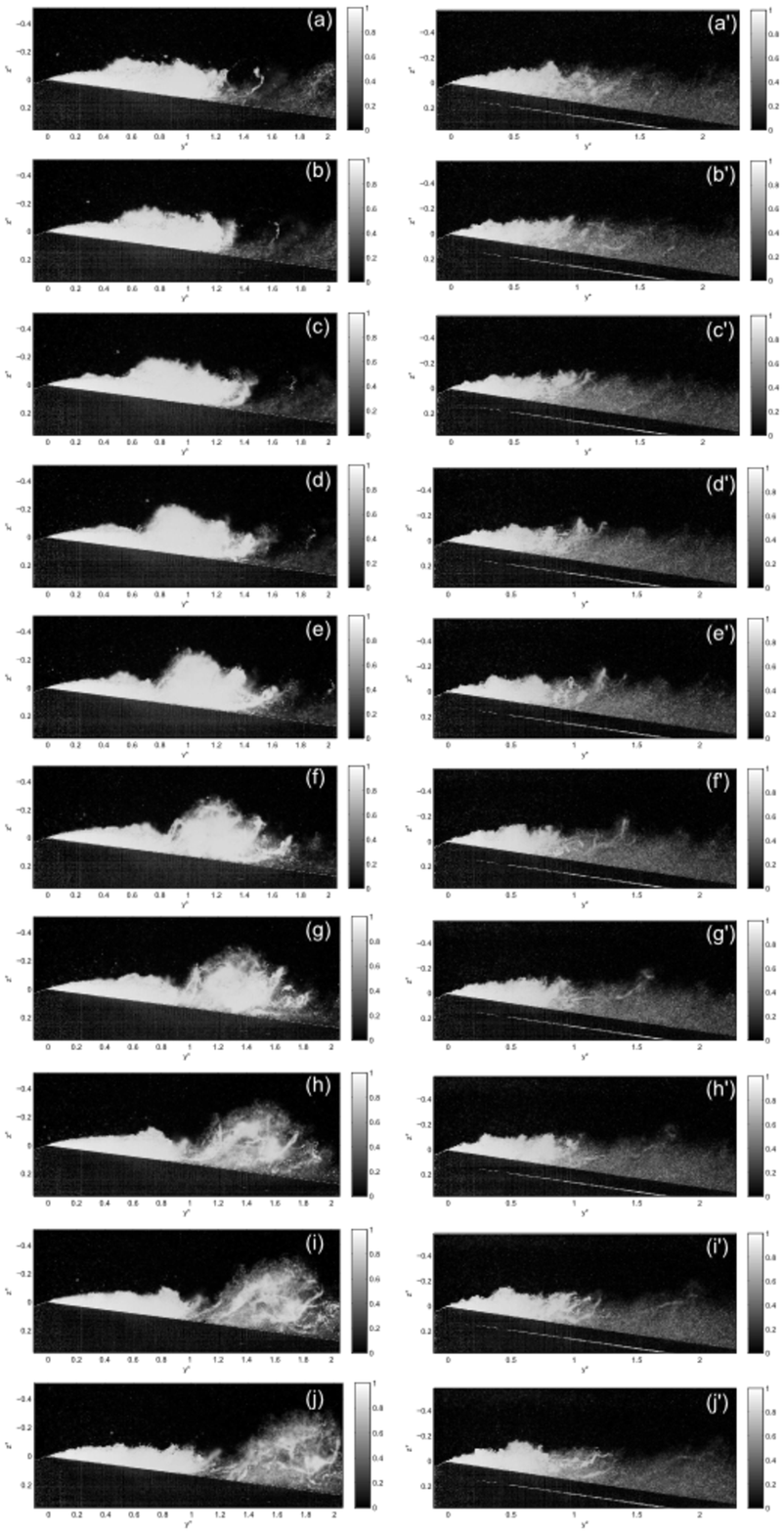,width=4.85in}}
\caption{Visualisation of (a-j) the cloud cavitation shedding for the smooth plate $0$ and (a'-j') the sheet cavity pulsation for the plate $8$, when $\mathit{\sigma=}1.17$ ($\mathrm{\Delta}\mathit{t}=2\,\mathrm{ms}$ between two images)}
\label{fig:Comp_plate8_lisse}
\end{figure*}

Plates $7$ and $8$ have a quite constant Strouhal number $0.05\leq \mathit{St_{L}} \leq 0.15$, which is characteristic of the sheet cavity regime, with a pulsating sheet cavity and no cloud cavitation shedding. Figure~\ref{fig:Comp_plate8_lisse} compares the sheet cavity evolutions at a same $\mathit{\sigma =}1.17$, for the reference smooth plate $0$ and the grooved plate $8$. It can be seen that in the case of the plate $8$ the sheet cavity pulses (Fig.~\ref{fig:Comp_plate8_lisse}(a)) without any shedding contrary to a smooth surface of the Venturi profile (Fig.\ref{fig:Comp_plate8_lisse}(b)). Then, the grooves depth $\mathit{h}$ controls the cavitation regime and can suppress the cloud cavitation for a large range of $\mathit{\sigma}$. Plates $7$ and $8$ are thus able to stabilize the sheet cavity and to reduce its length.It is thus possible to obtain a passive control by using specific geometrical parameters of grooves.

With the study of the sheet cavity height evolution (section~\ref{SheetCavitySize}), we can also determine the frequency $\mathit{f_{H}}$ of the sheet cavity height. This frequency $\mathit{f_{H}}$ and the frequency $\mathit{f_{L}}$ of the sheet cavity length evolution have been compared for plates $0$ to $6$. Results for plates $7$ and $8$ are not compared because for all the studied range of $\mathit{\sigma}$, there is no cloud cavitation shedding. For all grooved plates and for the smooth plate $0$, the curve $\mathit{f_{H}}=g\left(\mathit{f_{L}}\right)$ is linear so the frequency of the height evolution is linked to the length evolution frequency.

\section{Grooved surfaces effect on a non cavitating flow}\label{sec:NonCavitantFlow}

To discuss the reasons for the modifications of cavitating flows with the grooved surfaces, the velocity profiles are measured in the non cavitating flow, and more especially in the boundary layer, as it is the cavitation inception zone. Therefore, velocity measurements with Laser Doppler Velocimetry have been conducted.

\subsection{Laser Doppler Velocimetry measurements (LDV)}\label{LDV}

A one dimensional FlowExplorer Mini LDA system of Dantec Dynamics has been used to measure the longitudinal componant $\mathit{v_{y}}$ of the flow. This system consists in a factory-aligned and calibrated optical probe, with a focal lens of $300\,\mathrm{mm}$, and a signal processor. The optical head provides two laser beams with $25\,\mathrm{mW}$ power and a wave length $660\,\mathrm{nm}$. The measurement volume dimensions are $0.1\,\mathrm{mm}$ in diameter and $1\,\mathrm{mm}$ in length. The measurable velocity fluctuation is $0.002\%$ of the velocity range. This sytem has a high accuracy, as the calibration coefficient uncertainty is lower than $0.1\%$. In each measurement position, $10000$ samples are acquired, with a limiting time of acquisition fixed of $60~s$. The data rate is near $300\,\mathrm{Hz}$, while the validation rate is upper than $60\%$. For measurements downstream the Venturi throat, the optical head is inclined at $8^{\circ}$ in order to align the laser beams with the Venturi slope. In this way, measurements can be acquired very near the wall so velocity measurements can be used to analyze the effect of surface condition on the boundary layer of the flow. Then, the acquired velocity component is not the longitudinal velocity of the test section but the velocity component parrallel to the slope of the Venturi bottom wall. In the discussion of LDV results, this component is then called $\mathit{v'_{y}}$ in order to clarify explanations. 

On the other hand, velocity measurements presented on Fig.\ref{fig:VitesseAmont} were made with a horizontal optical head, so the component is really the longitudinal velocity $\mathit{v_{y}}$ in this case.

\subsection{Effects of the grooves on the velocity profile of the non cavitating flow}\label{NonCavitantFlow}

\begin{figure*}
\centerline{\psfig{figure=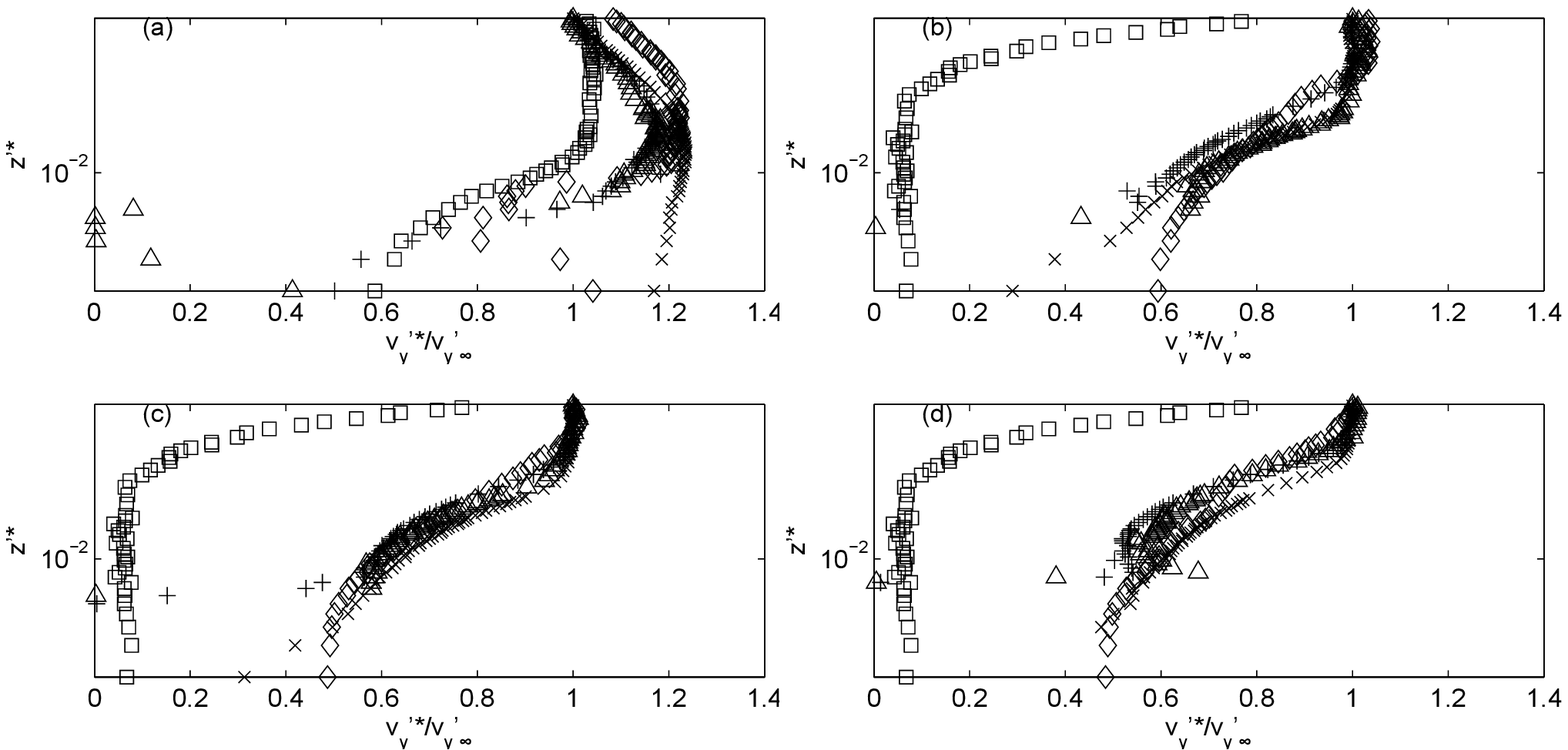,width=6.85in}}
  \caption{Longitudinal non-dimensional velocity profile $\mathit{v_{y}^{'*}/v'_{y_{\infty}}}$ of the non-cavitant flow, at (\textit{a}) $\mathit{y^{*}}=0$, (\textit{b}) $\mathit{y^{*}}=0.5$, (\textit{c}) $\mathit{y^{*}}=1$ and (\textit{d}) $\mathit{y^{*}}=2$. \textcolor{black}{$\times$}: plate $0$, $\diamondsuit$: plate $2$, \textcolor{black}{$+$}: plate $6$, \textcolor{black}{$\diamondsuit$}: plate 7, \textcolor{black}{$\Box$}: plate 8.}
\label{fig:ProfilsNonCavitantTousy}
\end{figure*}

Figure~\ref{fig:ProfilsNonCavitantTousy} shows undimensional velocity profiles $\mathit{v_{y}^{'*}=v'_{y}/v'_{\infty}}$, where $\mathit{v'_{y}}$  is the flow velocity component parallel to the bottom wall Venturi slope and $\mathit{v'_{\infty}}$ is the same velocity component measured far from the plate wall, at $\mathit{z^{'*}=z'/H_{throat}=}0.9$ (this zone is not shown on the figure). Velocity profiles are represented here at different distances from the throat, in $\mathit{y^{*}=}0.5$, $\mathit{y^{*}=}1$ and $\mathit{y^{*}=}2$. We can see on this Fig.\ref{fig:ProfilsNonCavitantTousy} that grooves modify the flow, near the wall, up to $\mathit{z^{'*}=}0.07$ for plates $2$, $6$ and $7$ and until $\mathit{z^{'*}=}0.3$ for plate $8$. This modification depends on the geometry of the grooves. For all values of $\mathit{y^{*}}$, excluding the plate $8$ for which the velocity profile is totally different, the velocity is always lower for the plate $6$, which is the plate with larger grooves and minimum number $\mathit{N}$ of grooves. On the other hand, for smaller grooves, with the plate $2$ for example ($\mathit{d}=1\,\mathrm{mm}$, $\mathit{h}=0.25\,\mathrm{mm}$ and $\mathit{e}=0.1\,\mathrm{mm}$) which has a large grooves number $\mathit{N=}124$, in $\mathit{y^{*}=}0.5$, the velocity profile is similar to the velocity profile of the reference plate $0$, without grooves. At station $\mathit{y^{*}}=1$ or $\mathit{y^{*}}=2$ the velocity for the plate $2$ becomes lower than for the reference plate $0$. The effect of plate $2$ on the grooves overlaps even the curve obtained with plate $6$. Then the effect of the grooves geometry disapears from $\mathit{y^{*}}=2$, but grooves still play the role of a brake for the flow. Very close to the wall, from $\mathit{z^{'*}}=0$ until $\mathit{z^{'*}}=0.02$, the velocity profile $\mathit{v_{y}^{'*}}$ for the plate $7$ is superimposed on the velocity profile of the plate $0$, when $0\leq \mathit{y^{'*}} \leq 2$. This feature explains why plate $7$ reduces sheet cavity length like plate $2$. If $\mathit{z^{'*}}>0.02$, then $\mathit{v_{y}^{'*}}$ is lower than other plates $2$ and $6$. On the other hand, the velocity profile obtained for the plate $8$ is totally different. $\mathit{v_{y}^{'*}}$ is very low from $\mathit{z^{'*}}=0.3$ for $0.5\leq \mathit{y^{*}} \leq 2$. This analysis permits to bring out the effect of structured roughness surface on the flow but also the importance of the geometric parameters of the grooves. The grooves wavelength is thus a crucial parameter to influence the sheet cavity dynamics, since plates $7$ and $8$ have the same grooves geometry but present different grooves wavelength.

\begin{figure*}
\centerline{\psfig{figure=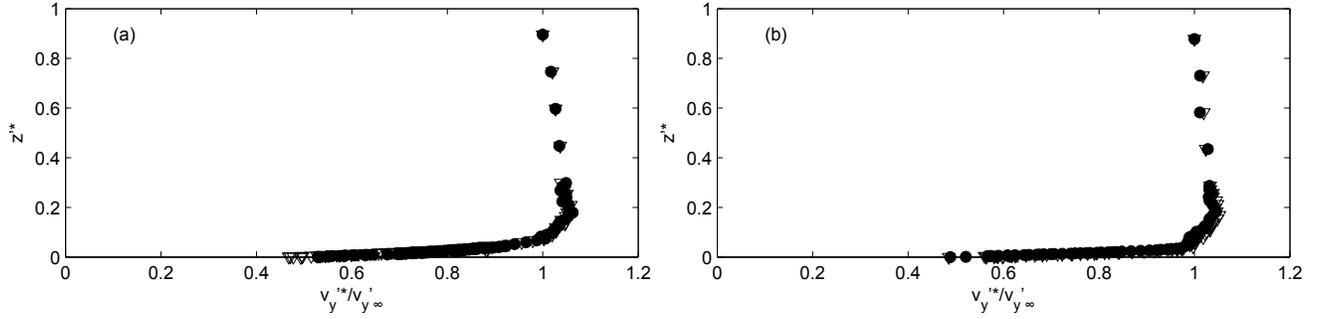,width=6.85in}}
  \caption{Longitudinal non-dimensional velocity profile $\mathit{v_{y}^{'*}/v'_{y_{\infty}}}$ of the non-cavitating flow, at $\mathit{y^{*}}=0.5$, for (\textit{a}) the grooved sheet $6$ and (\textit{b}) the grooved sheet $2$. $\bullet$: $x^{*}=0$, \textcolor{black}{$\triangle$}: $\mathit{x^{*}}=\frac{\mathit{\lambda}}{2}$}
\label{fig:ProfilNonCavitant}
\end{figure*}

In order to investigate in more details the effects of each groove on the velocity profile, two specific velocity profiles have been acquired for each plate: the first one is located in the middle of the test section wide, at $\mathit{x^{*}}=0$, above the recess of the groove, while the second one is measured at $\mathit{x^{*}}=\lambda/2$, above the edge of the groove. It can be observed (Fig.~\ref{fig:ProfilNonCavitant}) that the relative difference between the two curves is always lower than $3\%$. The effect of the grooves is thus not local but global, by the way it changes the flow instabilities.   

This non-cavitating flow analysis provides a basis for analysis of measurements performed in cavitating conditions. As it was reported previously, the use of plate $6$ results in a decrease of the velocity flow near the wall. The fluid encountered a larger surface to flow with larger grooves, so the velocity is considerably decreased. But with the analysis of the mean sheet cavity length, we can say that larger grooves decreases the cavity length. Moreover, with larger grooves, the re-entrant jet can travel further upstream before reaching the interface and cutting the sheet cavity, in order to extract a cloud cavitation. The immediate consequence is that the mean sheet cavity length should be smaller with large grooves. An hypothesis that can be proposed is that some grooves provide small recirculation zones that prevent the re-entrant jet to flow upstream, underneath the sheet cavity, as it is discussed in \cite{Coustols2001}, for the study of a non-cavitant flow boundary layer. Then the re-entrant jet is stopped by these recirculation zones. Not only re-entrant jet is crucial for the cavitation shedding but also side-entrant jets \cite{Koop2010}. The three-dimensional distribution of grooves and their geometry have of course effects on the side-entrant jets dynamics. \\

\section{Conclusions}

Effects of the surface condition of a Venturi profile have been investigated using visualisation, Laser Doppler Velocimetry, and pressure measurements. The aim of this study was to evaluate the importance of grooved surfaces provided by machining. Geometry of these grooved surfaces have been observed in order to identify crucial parameters. Results show that the depth of grooves is a determining factor. It is demonstrated that a grooves depth smaller than the viscous sublayer thickness has no effect on the sheet cavity length. Some plates lead to a cavitation instabilities changing: essentially plates $7$ and $8$, for which the sheet cavity length is reduced and the plate $6$ which increases this sheet cavity length. These three plates have a large depth $\mathit{h}\geq 1\,\mathrm{mm}$. If the depth is $\mathit{h}>\frac{\mathit{d}}{2}$, then the grooved plate is able to reduce the sheet cavity length. The study of the shedding frequency of cloud cavitation, by image processing highlighted two cavitation regimes: for small cavitation numbers ($1<\mathit{\sigma}<1.3$), cavitation is in an unstable regime, with large sheet cavity and large shedding of cloud cavitation. This regime is characterized by a Strouhal number $\mathit{St_{L}}=0.3$. The second cavitation regime is a stable regime, with $1.3<\mathit{\sigma}<1.9$, which presents sheet cavities with only oscillations in the downstream part (this oscillation being related to the blade passing frequency of the circulation pump used in the test rig). A large grooves depth $\mathit{h}$, can modify the sheet cavity regime. Plates $7$ and $8$, with $\mathit{h}=2\,\mathrm{mm}$, are examples of roughness surfaces that can suppress the cloud cavitation shedding in a large range of cavitation number $\mathit{\sigma}$. This study has thus demonstrated the feasability of a passive control of cavitation on a Venturi profile by modifying the surface condition of the bottom wall using distributed organized roughness.


%
% ...

\bibliographystyle{asmems4}

\bibliography{mybiblio}

\end{document}